\documentclass[10pt,english]{article}
\usepackage{times}
\usepackage[T1]{fontenc}
\usepackage[latin1]{inputenc}
\usepackage{amsmath}
\usepackage{amssymb}

\makeatletter

\providecommand{\tabularnewline}{\\}

\newcommand{\tensor}[1]{\stackrel{\leftrightarrow}{#1}}

\usepackage{slashed}

\usepackage{babel}
\makeatother
\begin{document}

\title{\textbf{CHARGES AND MASS SPECTRUM IN THE PISANO-PLEITEZ-FRAMPTON
3-3-1 GAUGE MODEL }}

\author{ION I. COT\u{A}ESCU and ADRIAN PALCU}

\date{\emph{Department of Theoretical and Computational Physics - West
University of Timi\c{s}oara, V. P\^{a}rvan Ave. 4, RO - 300223 Romania}}

\maketitle
\begin{abstract}
The Pisano-Pleitez-Frampton 3-3-1 model is revisited here within the
framework of the general method for solving gauge models with high
symmetries. This exact algebraical approach - proposed several years
ago by one of us - was designed to include a minimal Higgs mechanism
that spontaneously breaks the gauge symmetry up to the universal $U(1)_{em}$
electromagnetic one and, consequently, to supply the mass spectrum
and the couplings of the currents for all the particles in the model.
We prove in this paper that this powerful tool, when is applied to
the PPF 3-3-1 model, naturally recovers the whole Standard Model phenomenology
and, in addition, predicts - since a proper parametrization is employed
- viable results such as: (i) the exact expressions for the boson
and fermion masses, (ii) the couplings of the charged and neutral
currents and (iii) a plausible neutrino mass pattern. A generalized
Weinberg transformation is implemented, while the mixing between the
neutral bosons $Z$ and $Z^{\prime}$ is performed as a necessary
step by the method itself. Some phenomenological consequences are
also sketched, including the strange possibility that simultaneously
$m(Z)=m(Z^{\prime})$ and $m(W)=m(V)$ hold.

PACS numbers: 12.10.Dm; 12.60.Fr; 12.60.Cn; 14.60.Pq; 12.15. Mm;.

Key words: 3-3-1 gauge models, boson mass spectrum, neutrino masses
and mixings, neutral currents 
\end{abstract}

\section{Introduction}

Among the various extensions of the Standard Model (SM) that emerged
in the last decades in order to incorporate new phenomenology in the
electro-weak sector (such as neutrino oscillations, extra-neutral
bosons), or explain some features (such as mass hierarchy, fermion
families replication, CP-phase question, \emph{etc}), the well-known
Pisano-Pleitez-Frampton (PPF) model earned a wide reputation. It is
based on the gauge group $SU(3)_{C}\otimes SU(3)_{L}\otimes U(1)_{Y}$
(hereafter 3-3-1) that undergoes a spontaneously symmetry breakdown
(SSB) in two steps in order to provide with masses all the particles
in the model. The model was first proposed \cite{key-1,key-2} at
the beginning of the '90s and developed in the coming years with important
results regarding topics realated to the flavor changing neutral currents
(FCNC) \cite{key-3} - \cite{key-7} - including a proper GIM mechanism
for their suppression \cite{key-3} - , the CP-phase issue \cite{key-6}
- \cite{key-11}, the mass generation in the fermion sector \cite{key-12},
and the quest for an appropriate scalar sector \cite{key-13} - \cite{key-17}
for a realistic SSB. In order to make it a suitable gauge theory,
the 3-3-1 class of models has to be anomaly-free. A systematic approach
to the general case of the anomaly cancelation in 3-3-1 models can
be found in Refs. \cite{key-18,key-19}.

These different ways of phenomenologically investigating the 3-3-1
models seemed to explain only particular issues. Therefore, different
approximations were employed to solve certain troublesome aspects.
since a global view on the gauge models with high symmetries was still
lacking. This state of affairs called for an elegant and systematic
approach devoted to gauge theories with high symmetries in order to
make them able to supply general predictions, once a parameter set
is chosen from the very beginning of the calculus. Such an efficient
tool was proposed by Cot\u{a}escu in Ref. \cite{key-20} for the
general case of a theory with the electro-weak sector's symmetry given
by the gauge group $SU(n)_{L}\otimes U(1)_{Y}$ that undergoes a spontaneously
breakdown up to the universal electromagnetic one $U(1)_{em}$ in
one step only. For this purpose, the main parameters of such a theory
play the role of orthonormalization coefficients in the geometrized
scalar sector of the model, so that only one physical Higgs real field
finally survives the SSB. The method was successfully applied in a
recent series of papers \cite{key-21} - \cite{key-25} by Palcu in
the particular case of the 3-3-1 model with right-handed neutrinos,
a particular version of 3-3-1 models initially proposed in \cite{key-3},
and ever since championed by Long and his collaborators \cite{key-26}-\cite{key-33}.
The promising results given by the general method in that case encouraged
us to revisit the PPF model in order to respectively reveal its rich
phenomenology and embed the neutrino masses in it.

The general method is briefly reviewed in Sec.2 and subsecquently
is applied to the PPF 3-3-1 model. We prove that it supplies viable
results concerning both the boson mass spectrum and the neutral currents
of the model (Sec.3), and even is it able to generate fermion masses
(Sec.4) - including a suitable neutrino mass pattern - in accordance
with the available data \cite{key-34}. In Sec. 5 we give some conclusions
and phenomenological aspects of the results obtained, outlining particular
values of the main free parameter that can supply simultaneously $m(Z)=m(Z^{\prime})$
and $m(W)=m(V)$ which could, in turn, explain why those new bosons
were not yet experimentally discovered.

\section{The General Method of Solving Gauge Models}

In this section we recall the main results of the method of exactly
solving generalized $SU(n)_{L}\otimes U(1)_{Y}$ electro-weak gauge
models with a special type of Higgs mechanism proposed in Ref. \cite{key-20}.

\subsection{$SU(n)_{L}\otimes U(1)_{Y}$ electro-weak gauge models}

In our general approach, the basic piece involved in the gauge symmetry
is the group $SU(n)$. Its two fundamental irreducible unitary representations
(irreps) $\mathbf{n}$ and $\mathbf{n^{*}}$ play a crucial role in
constructing different classes of tensors of ranks $(r,s)$ as direct
products like $(\otimes{\textbf{n}})^{r}\otimes(\otimes{\textbf{n}}^{*})^{s}$.
These tensors have $r$ lower and $s$ upper indices for which we
reserve the notation, $i,j,k,\cdots=1,\cdots,n$. As usually, we denote
the irrep $\rho$ of $SU(n)$ by indicating its dimension, ${\mathbf{n}}_{\rho}$.
The $su(n)$ algebra can be parameterized in different ways, but here
it is convenient to use the hybrid basis of Ref. \cite{key-20} consisting
of $n-1$ diagonal generators of the Cartan subalgebra, $D_{\hat{i}}$,
labeled by indices $\hat{i},\hat{j},...$ ranging from $1$ to $n-1$,
and the generators $E_{j}^{i}=H_{j}^{i}/\sqrt{2}$, $i\not=j$, related
to the off-diagonal real generators $H_{j}^{i}$ \cite{key-35,key-36}.
This way the elements $\xi=D_{\hat{i}}\xi^{\hat{i}}+E_{j}^{i}\xi_{i}^{j}\in su(n)$
are now parameterized by $n-1$ real parameters, $\xi^{\hat{i}}$,
and by $n(n-1)/2$ $c$-number ones, $\xi_{j}^{i}=(\xi_{i}^{j})^{*}$,
for $i\not=j$. The advantage of this choice is that the parameters
$\xi_{j}^{i}$ can be directly associated to the $c$-number gauge
fields due to the factor $1/\sqrt{2}$ which gives their correct normalization.
In addition, this basis exhibit good trace orthogonality properties,
\begin{equation}
Tr(D_{\hat{i}}D_{\hat{j}})=\frac{1}{2}\delta_{\hat{i}\hat{j}},\quad Tr(D_{\hat{i}}E_{j}^{i})=0\,,\quad Tr(E_{j}^{i}E_{l}^{k})=\frac{1}{2}\delta_{l}^{i}\delta_{j}^{k}\,.\label{Eq.1}\end{equation}
 When we consider different irreps, $\rho$ of the $su(n)$ algebra
we denote $\xi^{\rho}=\rho(\xi)$ for each $\xi\in su(n)$ such that
the corresponding basis-generators of the irrep $\rho$ are $D_{\hat{i}}^{\rho}=\rho(D_{\hat{i}})$
and $E_{j}^{\rho\, i}=\rho(E_{j}^{i})$.

The $U(1)_{Y}$ transformations are nothing else but phase factor
multiplications. Therefore - since the coupling constants $g$ for
$SU(n)_{L}$ and $g^{\prime}$ for the $U(1)_{Y}$ are assinged -
the transformation of the fermion tensor $L^{\rho}$ with respect
to the gauge group of the theory reads \begin{equation}
L^{\rho}\rightarrow U(\xi^{0},\xi)L^{\rho}=e^{-i(g\xi^{\rho}+g^{\prime}y_{ch}\xi^{0})}L^{\rho}\label{Eq.2}\end{equation}
 where $\xi=\in su(n)$ and $y_{ch}$ is the chiral hypercharge defining
the irrep of the $U(1)_{Y}$ group parametrized by $\xi^{0}$. For
simplicity, the general method deals with the character $y=y_{ch}g^{\prime}/g$
instead of the chiral hypercharge $y_{ch}$, but this mathematical
artifice does not affect in any way the results. Therefore, the irreps
of the whole gauge group $SU(n)_{L}\otimes U(1)_{Y}$ are uniquely
detemined by indicating the dimension of the $SU(n)$ tensor and its
character $y$ as $\rho=({\textbf{n}}_{\rho},y_{\rho})$.

In general, the spinor sector of our models has at least a part (usually
the leptonic one) which is put in pure left form using the charge
conjugation. Consequently this includes only left components, $L=\sum_{\rho}\oplus L^{\rho}$,
that transform according to an arbitrary reducible representation
of the gauge group. The Lagrangian density of this part of the spinor
sector may have the form \begin{equation}
{\mathcal{L}}_{S_{0}}=\frac{i}{2}\sum_{\rho}\overline{L^{\rho}}\tensor{\not\!\partial}L^{\rho}-\frac{1}{2}\sum_{\rho\rho'}\left(\overline{L^{\rho}}\chi^{\rho\rho^{\prime}}(L^{\rho^{\prime}})^{c}+h.c.\right).\label{Eq.3}\end{equation}
 Bearing in mind that each left-handed multiplet transforms as $L^{\rho}\rightarrow U^{\rho}(\xi^{0},\xi)L^{\rho}$
we understand that ${\mathcal{L}}_{S_{0}}$ remains invariant under
the global $SU(n)_{L}\otimes U(1)_{Y}$ transformations if the blocks
$\chi^{\rho\rho'}$ transform like $\chi^{\rho\rho^{\prime}}\rightarrow U^{\rho}(\xi^{0},\xi)\chi^{\rho\rho^{\prime}}(U^{\rho^{\prime}}(\xi^{0},\xi))^{T}$,
according to the representations $({\textbf{n}}_{\rho}\otimes{\textbf{n}}_{\rho'},y_{\rho}+y_{\rho'})$
which generally are reducible. These blocks will give rise to the
Yukawa couplings of the fermions with the Higgs fields. The spinor
sector is coupled to the standard Yang-Mills sector constructed in
usual manner by gauging the $SU(n)_{L}\otimes U(1)_{Y}$ symmetry
\cite{key-20}. To this end we introduce the gauge fields $A_{\mu}^{0}=(A_{\mu}^{0})^{*}$
and $A_{\mu}=A_{\mu}^{+}\in su(n)$. Furthermore, the ordinary derivatives
are replaced in Eq. (\ref{Eq.3}) by the covariant ones, defined as
$D_{\mu}L^{\rho}=\partial_{\mu}L^{\rho}-ig(A_{\mu}^{\rho}+y_{\rho}A_{\mu}^{0})L^{\rho}$
thus arriving to the interaction terms of the spinor sector.

The Higgs sector, organized as the so called minimal Higgs mechanism
\cite{key-20}, is able to produce maximal effects but with only one
remaining Higgs neutral field, just as in SM. This sector consists
of $n$ Higgs multiplets $\phi^{(1)}$, $\phi^{(2)}$, ... $\phi^{(n)}$
satisfying the orthogonality condition $\phi^{(i)+}\phi^{(j)}=\phi^{2}\delta_{ij}$
in order to eliminate the unwanted Goldstone bosons that could survive
the SSB. $\phi$ is a gauge-invariant real scalar field while the
Higgs multiplets $\phi^{(i)}$ transform according to the irreps $({\textbf{n}},y^{(i)})$
whose characters $y^{(i)}$ are arbitrary numbers that can be organized
into the diagonal matrix \begin{equation}
Y={\textrm{diag}}\left(y^{(1)},y^{(2)},\cdots,y^{(n)}\right)\,.\label{Eq.4}\end{equation}
 The Higgs sector is constructed by resorting to the parameter matrix
\begin{equation}
\eta={\textrm{diag}}\left(\eta{}^{(1)},\eta{}^{(2)},...,\eta{}^{(n)}\right)\label{Eq.5}\end{equation}
 with the property ${\textrm{Tr}}(\eta^{2})=1-\eta_{0}^{2}$. It will
play the role of the metric in the kinetic part of the Higgs Lagrangian
density which reads \begin{equation}
\mathcal{L}_{H}=\frac{1}{2}\eta_{0}^{2}\partial_{\mu}\phi\partial^{\mu}\phi+\frac{1}{2}\sum_{i=1}^{n}\left(\eta{}^{(i)}\right)^{2}\left(D_{\mu}\phi^{(i)}\right)^{+}\left(D^{\mu}\phi^{(i)}\right)-V(\phi)\label{Eq.6}\end{equation}
 where $D_{\mu}\phi^{(i)}=\partial_{\mu}\phi^{(i)}-ig(A_{\mu}+y^{(i)}A_{\mu}^{0})\phi^{(i)}$
are the covariant derivatives of the model and $V(\phi)$ is the scalar
potential generating the SSB of the gauge symmetry \cite{key-20}.
This is assumed to have an absolute minimum for $\phi=\langle\phi\rangle\not=0$
that is, $\phi=\langle\phi\rangle+\sigma$ where $\sigma$ is the
unique surviving physical Higgs field. Therefore, one can always define
the unitary gauge where the Higgs multiplets, $\hat{\phi}^{(i)}$,
have the components \begin{equation}
\hat{\phi}_{k}^{(i)}=\delta_{ik}\phi=\delta_{ik}(\langle\phi\rangle+\sigma)\,.\label{Eq.7}\end{equation}

This will be of great importance when the fermion masses will be computed,
due to the fact that the fermion mass terms - provided by Eq. (\ref{Eq.3})
via this minimal Higgs mechanism (mHm) - exhibit the Yukawa traditional
form only when the theory is boosted towards the unitary gauge.

\subsection{Neutral bosons}

A crucial goal is now to find the physical neutral bosons with well-defined
properties. This must start with the separation of the electromagnetic
potential $A_{\mu}^{em}$ corresponding to the surviving $U(1)_{em}$
symmetry. We have shown that the one-dimensional subspace of the parameters
$\xi^{em}$ associated to this symmetry assumes a particular direction
in the parameter space $\lbrace\xi^{0},\xi^{\hat{i}}\rbrace$ of the
whole Cartan subalgebra. This is uniquely determined by the $n-1$
- dimensional unit vector $\nu$ and the angle $\theta$ giving the
subspace equations $\xi^{0}=\xi^{em}\cos\theta$ and $\xi^{\hat{i}}=\nu_{\hat{i}}\xi^{em}\sin\theta$.
On the other hand, since the Higgs multiplets in unitary gauge remain
invariant under $U(1)_{em}$ transformations, we must impose the obvious
condition $D_{\hat{i}}\xi^{\hat{i}}+Y\xi^{0}=0$ which yields \begin{equation}
Y=-D_{\hat{i}}\nu^{\hat{i}}\tan\theta\equiv-(D\cdot\nu)\tan\theta\,.\label{Eq.8}\end{equation}
 In other words, the new parameters $(\nu,\theta)$ determine all
the characters $y^{(i)}$ of the irreps of the Higgs multiplets. For
this reason these will be considered the principal parameters of the
model and therefore one deals with $\theta$ and $\nu$ (which has
$n-2$ independent components) instead of $n-1$ parameters $y^{(i)}$.

Under these circumstances, the generating mass term \begin{equation}
\frac{g^{2}}{2}\langle\phi\rangle^{2}Tr\left[\left(A_{\mu}+YA_{\mu}^{0}\right)\eta^{2}\left(A^{\mu}+YA^{0\mu}\right)\right]\,,\label{Eq.9}\end{equation}
 depends now on the parameters $\theta$ and $\nu_{\hat{i}}$. The
neutral bosons in Eq. \ref{Eq.9} being the electromagnetic field
$A_{\mu}^{em}$ and the $n-1$ new ones, $A_{\mu}^{'\hat{i}}$, which
are the diagonal bosons remaining after the separation of the electromagnetic
potential \cite{key-20}.

This term straightforwardly gives rise to the masses of the non-diagonal
gauge bosons \begin{equation}
M_{i}^{j}=\frac{1}{2}g\left\langle \phi\right\rangle \sqrt{\left[\left(\eta^{(i)}\right)^{2}+\left(\eta^{(j)}\right)^{2}\right]}\,,\label{Eq.10}\end{equation}
 while the masses of the neutral bosons $A_{\mu}^{'\hat{i}}$ have
to be calculated by diagonalizing the matrix \begin{equation}
(M^{2})_{\hat{i}\hat{j}}=\langle\phi\rangle^{2}Tr(B_{\hat{i}}B_{\hat{j}})\label{Eq.11}\end{equation}
 where \begin{equation}
B_{\hat{i}}=g\left(D_{\hat{i}}+\nu_{\hat{i}}(D\cdot\nu)\frac{1-\cos\theta}{\cos\theta}\right)\eta,\label{Eq.12}\end{equation}
 As it was expected, $A_{\mu}^{em}$ does not appear in the mass term
and, consequently, it remains massless. The other neutral gauge fields
${A'}_{\mu}^{\hat{i}}$ have the non-diagonal mass matrix (\ref{Eq.11}).
This can be brought in diagonal form with the help of a new $SO(n-1)$
transformation, $A_{\mu}^{'\hat{i}}=\omega_{\cdot\;\hat{j}}^{\hat{i}\;\cdot}Z_{\mu}^{\hat{j}}$
, which leads to the physical neutral bosons $Z_{\mu}^{\hat{i}}$
with well-defined masses. Performing this $SO(n-1)$ transformation
the physical neutral bosons are completely determined. The transformation
\begin{eqnarray}
A_{\mu}^{0} & = & A_{\mu}^{em}\cos\theta-\nu_{\hat{i}}\omega_{\cdot\;\hat{j}}^{\hat{i}\;\cdot}Z_{\mu}^{\hat{j}}\sin\theta,\nonumber \\
A_{\mu}^{\hat{k}} & = & \nu^{\hat{k}}A_{\mu}^{em}\sin\theta+\left(\delta_{\hat{i}}^{\hat{k}}-\nu^{\hat{k}}\nu_{\hat{i}}(1-\cos\theta)\right)\omega_{\cdot\;\hat{j}}^{\hat{i}\;\cdot}Z_{\mu}^{\hat{j}}.\label{Eq.13}\end{eqnarray}
 which switches from the original diagonal gauge fields, $(A_{\mu}^{0},A_{\mu}^{\hat{i}})$
to the physical ones, $(A_{\mu}^{em},Z_{\mu}^{\hat{i}})$ is called
the generalized Weinberg transformation (gWt).

The nest step is to identify the charges of the particles with the
coupling coefficients of the currents with respect to the above determined
physical bosons. Thus, we find that the spinor multiplet $L^{\rho}$
(of the irrep $\rho$) has the following electric charge matrix \begin{equation}
Q^{\rho}=g\left[(D^{\rho}\cdot\nu)\sin\theta+y_{\rho}\cos\theta\right],\label{Eq.14}\end{equation}
 and the $n-1$ neutral charge matrices \begin{equation}
Q^{\rho}(Z^{\hat{i}})=g\left[D_{\hat{k}}^{\rho}-\nu_{\hat{k}}(D^{\rho}\cdot\nu)(1-\cos\theta)-y_{\rho}\nu_{\hat{k}}\sin\theta\right]\omega_{\cdot\;\hat{i}}^{\hat{k}\;\cdot}\label{Eq.15}\end{equation}
 corresponding to the $n-1$ neutral physical fields, $Z_{\mu}^{\hat{i}}$.
All the other gauge fields, namely the charged bosons $A_{j\mu}^{i}$,
have the same coupling, $g/\sqrt{2}$, to the fermion multiplets.

At this point one can change again the parametrization by using the
electrical charges $q_{i}$ of the fundamental multiplet $({\textbf{n}},0)$
given by \begin{equation}
Q\equiv{\textrm{diag}}(q_{1},q_{2},\cdots,q_{n})=g(D\cdot\nu)\sin\theta\,,\label{Eq.16}\end{equation}
 instead of the parameters $(g,\nu_{\hat{i}})$ but keeping the angle
$\theta$ as the principal parameter of the model in order to remain
in the spirit of the SM. This way $g$ and $\nu_{\hat{i}}$ have to
be expressed in terms of $q_{i}$ using the formulas $g\nu_{\hat{i}}\sin\theta=2{\textrm{Tr}}(D_{\hat{i}}Q)$
and ${g}^{2}\sin^{2}\theta=2{\textrm{Tr}}(Q^{2})$. Moreover, the
matrix (\ref{Eq.4}) can be written now as $Y=-Q\tan\theta/\sqrt{2{\textrm{Tr}}(Q^{2})}$.
Finally we have to replace $y_{\rho}$ with $y_{ch}^{\rho}(g^{\prime}/g)$
in order to deal with the veritable chiral character of $U(1)_{Y}$.
The quantity $y_{ch}$ becomes the usual chiral hypercharge if we
take \begin{equation}
g'=e\, g\,\frac{\tan\theta}{\sqrt{2{\textrm{Tr}}(Q^{2})}}\,,\label{Eq.17}\end{equation}
 where $e$ is the elementary electric charge. This supplies at the
same time the correct relation between the two couplings $g$ and
$g'$ - once the $\theta$- angle is given as a function of the $\theta_{W}$
from SM - without resorting to any other supplemental condition. Particularly,
with this assignment the chiral hypercharges of the Higgs multiplets
take the simpler form $Y_{ch}=-Q/e$.

\section{The Pisano-Pleitez-Frampton 3-3-1 Model Revisited}

The general method - constructed in Ref. \cite{key-20} and briefly
presented in the above section - is based on the following assumptions
in order to give viable results when it is applied to concrete models:

({\small I}) the spinor sector must be put (at least partially) in
pure left form using the charge conjugation (see for details Appendix
B in Ref. \cite{key-20})

({\small II}) a minimal Higgs mechanism with arbitrary parameters
$(\eta_{0},\eta)$ satisfying the condition ${\textrm{Tr}}(\eta^{2})=1-\eta_{0}^{2}$
and giving rise to traditional Yukawa couplings in unitary gauge is
employed

({\small III}) the coupling constant, $g$, is the same with the first
one of the SM

({\small IV}) at least one $Z$-like boson should satisfy the mass
condition $m_{Z}=m_{W}/\cos\theta_{W}$ established in the SM and
experimentally confirmed.

Bearing in mind all these necessary ingredients, we proceed to solving
the particular 3-3-1 model of PPF \cite{key-1,key-2} by imposing
from the very beginning the set of parameters we will work with.

\subsection{The structure of the model}

In what follows we denote the irreps of the electro-weak model under
consideration here by $\rho=({\textbf{n}}_{\rho},y_{ch}^{\rho})$
indicating the veritable chiral hypercharge $y_{ch}$ instead of $y$.
Therefore, the multiplets of the PPF 3-3-1 model will be denoted by
$({\textbf{n}}_{color},{\textbf{n}}_{\rho},y_{ch}^{\rho})$. With
this notation the irreps of the spinor sector are:

\textbf{Lepton families}\begin{equation}
\begin{array}{ccccc}
f_{\alpha L}=\left(\begin{array}{c}
e_{\alpha}^{c}\\
e_{\alpha}\\
\nu_{\alpha}\end{array}\right)_{L}\sim(\mathbf{1,3},0) &  &  &  & \left(e_{\alpha L}\right)^{c}\sim(\mathbf{1},\mathbf{1},-1)\end{array}\label{Eq.18}\end{equation}

\textbf{Quark families}\begin{equation}
\begin{array}{ccc}
Q_{iL}=\left(\begin{array}{c}
J_{i}\\
u_{i}\\
d_{i}\end{array}\right)_{L}\sim(\mathbf{3,3^{*}},-1/3) &  & Q_{3L}=\left(\begin{array}{c}
J_{3}\\
-b\\
t\end{array}\right)_{L}\sim(\mathbf{3},\mathbf{3},+2/3)\end{array}\label{Eq.19}\end{equation}
\begin{equation}
\begin{array}{ccc}
(b_{L})^{c},(d_{iL})^{c}\sim(\mathbf{3},\mathbf{1},-1/3) &  & (t_{L})^{c},(u_{iL})^{c}\sim(\mathbf{3},\mathbf{1},+2/3)\end{array}\label{Eq.20}\end{equation}
\begin{equation}
\begin{array}{ccccccccc}
(J_{3L})^{c}\sim(\mathbf{3,1},+5/3) &  &  &  &  &  &  &  & (J_{iL})^{c}\sim(\mathbf{3,1},-4/3)\end{array}\label{Eq.21}\end{equation}
 with $\alpha=1,2,3$ and $i=1,2$. In the representations presented
above we assumed, like in majority of the papers in the literature,
that the third generation of quarks transforms differently from the
other two ones. This could explain the unusual heavy masses of the
third generation of quarks, and especially the uncommon properties
of the top quark. The capital letters $J$ denote the exotic quarks
included in each family.

With this assignment the fermion families cancel all the axial anomalies
by just an interplay between them, although each family remains anomalous
by itself. Thus, the renormalization criteria are fulfilled and the
method is validated once more.

Note that one can add at any time sterile neutrinos - \emph{i.e.}
right-handed neutrinos $\nu_{\alpha R}\sim(\mathbf{1,1},0)$ - that
could pair in the neutrino sector of the Lagrangian density with left-handed
ones in order to generate eventually tiny Dirac or Majorana masses
by means of an adequate see-saw mechanism. These sterile neutrinos
do not affect anyhow the anomaly cancelation, since all their charges
are zero. Moreover, their number is not restricted by the number of
flavors in the model

Subsequently, we will use the standard generators $T_{a}=\lambda_{a}/2$
of the $su(3)$ algebra connected to the usual Gell-Mann matrices
which are differently displayed from those of Ref. \cite{key-20}.
So, the Hermitian diagonal generators of the Cartan subalgebra are
\begin{equation}
D_{1}=T_{3}=\frac{1}{2}{\textrm{diag}}(1,-1,0)\,,\quad D_{2}=T_{8}=\frac{1}{2\sqrt{3}}\,{\textrm{diag}}(1,1,-2)\,.\label{Eq22}\end{equation}
 In this basis the gauge fields are $A_{\mu}^{0}$ and $A_{\mu}\in su(3)$
that is \begin{equation}
A_{\mu}=\frac{1}{2}\left(\begin{array}{ccc}
A_{\mu}^{3}+A_{\mu}^{8}/\sqrt{3} & \sqrt{2}U_{\mu} & \sqrt{2}W_{\mu}\\
\sqrt{2}U_{\mu}^{*} & -A_{\mu}^{3}+A_{\mu}^{8}/\sqrt{3} & \sqrt{2}V_{\mu}\\
\sqrt{2}W_{\mu}^{*} & \sqrt{2}V_{\mu}^{*} & -2A_{\mu}^{8}/\sqrt{3}\end{array}\right),\label{Eq.23}\end{equation}
 Apart from the charged Weinberg bosons, $W$, there are new charged
bosons, $V$ and $U$, among them $U$ is doubly charged - the so
called ''bilepton'' - coupling different chiral states of the same
charged lepton.

For our purpose it is convenient exploit the parametrization based
on the $\theta$ angle and the electric charges of the lepton multiplet.
The latter are supposed to be $Q=e\,{\textrm{diag}}\,(1,-1,0)$. On
the other hand, Eq. (\ref{Eq.14}) allows us to identify $\nu=(1,0)$
and $g\sin\theta=2e$. As long the SM condition $e=g\sin\theta_{W}$
holds, one obtains $\sin\theta=2\sin\theta_{W}$. It remains to observe
that we have $Y_{ch}=-Q/e={\textrm{diag}}(-1,1,0)$ which means that
the irreps of the Higgs sector are $\phi^{(1)}\sim({\textbf{3}},-1)$,
$\phi^{(2)}\sim({\textbf{3}},1)$ and $\phi^{(3)}\sim({\textbf{3}},0)$.
Note that the Higgs components in unitary gauge satisfy Eq. (\ref{Eq.7})
only if this numeration of the Higgs multiplets is kept.

\subsection{Boson mass spectrum}

The masses of both the neutral and charged bosons depend on the choice
of the matrix $\eta$ whose components are free parameters. Here it
is convenient to assume the following matrix \begin{equation}
\eta^{2}=(1-\eta_{0}^{2}){\textrm{diag}}\left(1-a,\frac{a+b}{2},\frac{a-b}{2}\right)\label{Eq.24}\end{equation}
 where, for the moment, $a$ and $b$ are arbitrary non-vanishing
real parameters. Obviously, $\eta_{0},a\in[0,1)$. Note that with
this parameter choice the condition (II) is accomplished. Under these
circumstances, the mass matrix of the neutral bosons Eq. (\ref{Eq.11})
reads \begin{equation}
M^{2}=m^{2}\left(\begin{array}{cc}
\frac{{\textstyle 1}}{{\textstyle \cos^{2}\theta}}\left(1-\frac{{\textstyle 1}}{{\textstyle 2}}a+\frac{{\textstyle 1}}{{\textstyle 2}}b\right) & \frac{{\textstyle 1}}{{\textstyle \sqrt{3}\cos\theta}}\left(1-\frac{{\textstyle 3}}{{\textstyle 2}}a-\frac{{\textstyle 1}}{{\textstyle 2}}b\right)\\
\frac{{\textstyle 1}}{{\textstyle \sqrt{3}\cos\theta}}\left(1-\frac{{\textstyle 3}}{{\textstyle 2}}a-\frac{{\textstyle 1}}{{\textstyle 2}}b\right) & \frac{{\textstyle 1}}{{\textstyle 3}}+\frac{{\textstyle 1}}{{\textstyle 2}}a-\frac{{\textstyle 1}}{{\textstyle 2}}b\end{array}\right)\label{Eq.25}\end{equation}
 with $m^{2}=g^{2}\left\langle \phi\right\rangle ^{2}(1-\eta_{0}^{2})/4$.
Let us observe that the condition (IV) is fulfilled if and only if
$b/a=-3\tan^{2}\theta_{W}$. That is, one remains with only one parameter
- say $a$. In addition, there are terms which become singular for
$\cos\theta=0$ which corresponds to the value $\sin^{2}\theta_{W}=1/4$.
On this reason the Weinberg angle is restricted in this particular
model to values for $\sin^{2}\theta_{W}$ less than $1/4$, which
is in good accord to experimental measurements on it \cite{key-34}.

The mass spectrum of the gauge bosons (without insisting on the computing
details) looks like \begin{eqnarray}
{m}^{2}(W) & = & {m}^{2}(Z)\cos^{2}\theta_{W}=m^{2}a,\nonumber \\
{m}^{2}(Z') & = & \frac{m^{2}}{1-4\sin^{2}\theta_{W}}\left\{ \frac{4}{3}\cos^{2}\theta_{W}-a\left[1-(1-4\sin^{2}\theta_{W})\tan^{2}\theta_{W}\right]\right\} \,,\nonumber \\
{m}^{2}(V) & = & m^{2}\left[1-\frac{a}{2}(1-3\tan^{2}\theta_{W})\right],\label{Eq.26}\\
{m}^{2}(U) & = & m^{2}\left[1-\frac{a}{2}(1+3\tan^{2}\theta_{W})\right]\,,\nonumber \end{eqnarray}
 Obviously, $Z$ is the neutral boson of the SM, while $Z'$ is the
new neutral boson of this model.

The mass scale is now just a matter of tuning the parameter $a$ in
accordance with the possible values for $\left\langle \phi\right\rangle $.
However, this mass spectrum exhibits a very strange feature. For the
particular value \begin{equation}
a=a_{c}=\frac{2\cos^{2}\theta_{W}}{3(1-2\sin^{2}\theta_{W})}\label{Eq.27}\end{equation}
 a \emph{critical point} arises. At that very value the following
equalities $m(Z)=m(Z^{\prime})$ and $m(W)=m(V)$ are simultaneously
fulfilled, while the bilepton mass becomes \begin{equation}
m^{2}(U)=m^{2}(Z)(1-3\sin^{2}\theta_{W}).\label{Eq.28}\end{equation}
 Numerically speaking if one inserts $\sin^{2}\theta_{W}\sim0.223$
in Eq. \ref{Eq.27} then one gets that the critical point corresponds
to $a_{c}\sim0.934$ and $m(U)\sim30\, GeV/c^{2}$. This phenomenon
could give a plausible explanation for why the new bosons were not
yet discovered and precisely weighted in the laboratory. However,
although the data \cite{key-34} suggest $m(Z)<m(Z^{\prime})$ the
possibility outlined above is not definitely ruled out unless an experimental
argument is invoked. We are confident that this issue will be elucidated
in the near future at LHC, when a precise experimental measurement
of the masses of these new bosons predicted by the 3-3-1 theory will
be available.

\subsection{Electric and neutral charges}

In the PPF 3-3-1 model under consideration here, assuming the versor
choice $\nu=(1,0)$, we obtain the generalized Wienberg transformation
which was designed to reach the physical basis $(A^{em},Z,Z^{\prime})$
of the neutral bosons of the model. This reads \begin{eqnarray}
A_{\mu}^{0} & = & A_{\mu}^{em}\cos\theta-\left(\omega_{\cdot1}^{1\cdot}Z_{\mu}^{\prime}+\omega_{\cdot2}^{1\cdot}Z_{\mu}\right)\sin\theta\nonumber \\
A_{\mu}^{3} & = & A_{\mu}^{em}\sin\theta+\left(\omega_{\cdot1}^{1\cdot}Z_{\mu}^{\prime}+\omega_{\cdot2}^{1\cdot}Z_{\mu}\right)\cos\theta\label{Eq.29}\\
A_{\mu}^{8} & = & \omega_{\cdot1}^{2\cdot}Z_{\mu}^{\prime}+\omega_{\cdot2}^{2\cdot}Z_{\mu}\nonumber \end{eqnarray}
 where $\omega$ acting as the required $SO(2)$ rotation. Its components
\begin{equation}
\omega_{\cdot1}^{1\cdot}=\omega_{\cdot2}^{2\cdot}=-\frac{\sqrt{3}}{2\cos\theta_{W}}\,,\quad\omega_{\cdot2}^{1\cdot}=-\omega_{\cdot1}^{2\cdot}=\frac{1}{2}\sqrt{1-3\tan^{2}\theta_{W}}\,,\label{Eq.30}\end{equation}
 ensure the diagonal form of the matrix (\ref{Eq.25}). In order to
recover all the results of SM and those of Ref. \cite{key-20} (up
to sign) the obvious identification has to be performed: $Z^{2}=Z$
and $Z^{1}=Z^{\prime}$. It is worth observing that at the critical
point, $a=a_{c}$, the matrix (\ref{Eq.25}) becomes proportional
with the unit matrix ${\textbf{1}}_{2\times2}$ so that the $\omega$-rotation
can be arbitrarily chosen, offering thus a supplementary degree of
freedom in defining $Z$-bosons. However, in order to avoid here a
digression on this subject, we restrict ourselves to keep the rotation
(\ref{Eq.30}) at the critical point too, following to discuss about
it elsewhere.

All the needed ingredients are now available in order to express the
content of the gauge sector, $A_{\mu}^{\rho}+y_{ch}^{\rho}A_{\mu}^{0}g'/g$,
in terms of physical neutral bosons ($A^{em}$, $Z$, $Z^{\prime}$)
as well as the charged ones of Eq. (\ref{Eq.23}), namely $(W^{\pm},V^{\pm},U^{2\pm})$.
The latter charged fields couple the currents of the spinor multiplets
$L^{\rho}$ through the coupling constant $g=e/\sin\theta_{W}$, while
from Eq. (\ref{Eq.17}) straightforwardly results \begin{equation}
g^{\prime}=g\frac{\sin\theta_{W}}{\sqrt{1-4\sin^{2}\theta_{W}}}\,.\label{Eq.31}\end{equation}
 Hereby we have to obtain the correct electric charges of the fermion
irreps and, subsequently, the expected neutral charges for all the
particles in the theory. In what follows all these coupling coefficients
will be written in units of elementary electric charge, $e$. The
electric charges of the components of a multiplet obeying the irrep
$\rho$ read \begin{equation}
Q^{\rho}(A^{em})=2T_{3}^{\rho}+y_{ch}^{\rho}\,,\label{Eq.32}\end{equation}
 while the neutral charges corresponding to the bosons $Z$ and $Z^{\prime}$
are \begin{eqnarray}
Q^{\rho}(Z) & = & \frac{1}{\sin2\theta_{W}}\left[T_{3}^{\rho}\left(1-4\sin^{2}\theta_{W}\right)-T_{8}^{\rho}\sqrt{3}-2y_{ch}^{\rho}\sin^{2}\theta_{W}\right]\,,\label{Eq.33}\\
Q^{\rho}(Z^{\prime}) & = & -\frac{\sqrt{1-4\sin^{2}\theta_{W}}}{\sin2\theta_{W}}\left(T_{3}^{\rho}\sqrt{3}+T_{8}^{\rho}-y_{ch}^{\rho}\frac{2\sin^{2}\theta_{W}}{1-4\sin^{2}\theta_{W}}\sqrt{3}\right)\,.\label{Eq.34}\end{eqnarray}
 It is remarkable that all the coupling coefficients of this model
are independent of the parameter $a$ responsible for the boson mass
spectrum. Computing the concrete values of these coefficients for
all the fermion multiplets is presented in detail in Appendix and
the results are displayed in Table.

\section{Fermion Masses}

Generating fermion masses is one of the most stringent issues in particle
physics. This question is addressed in this section within the PPF
3-3-1 model, assuming that the technique of the {}``classical''
Yukawa terms worked very well at SM level, although their couplings
remained unrestricted parameters on theoretical ground. These values
are exclusively determined by experimental reasons.

\subsection{Quark masses}

For all the quarks involved in the PPF 3-3-1 model, the traditional
Yukawa couplings seem to be sufficient in order to supply their desired
masses. That is - with the assignment of the Sec. 3.1. for the representations
in the fermion and scalar sectors - one has the following terms in
the quark mass sector: \begin{eqnarray}
 &  & G_{u}\bar{Q}_{1L}\phi^{(2)+}u_{R}+G_{c}\bar{Q}_{2L}\phi^{(2)+}c_{R}+H.c.\label{Eq.35}\\
 &  & G_{d}\bar{Q}_{1L}\phi^{(3)+}d_{R}+G_{s}\bar{Q}_{2L}\phi^{(3)+}s_{R}+H.c.\label{Eq.36}\\
 &  & G_{t}\bar{Q}_{3L}\phi^{(3)}t_{R}+G_{b}\bar{Q}_{3L}\phi^{(2)}b_{R}+H.c.\label{Eq.37}\\
 &  & G_{1}\bar{Q}_{1L}\phi^{(1)+}J_{iR}+G_{2}\bar{Q}_{2L}\phi^{(1)+}J_{iR}H.c\label{Eq.38}\\
 &  & G_{1}\bar{Q}_{3L}\phi^{(1)}J_{3R}+H.c\label{Eq.39}\end{eqnarray}
 These terms are assumed to undergo necessary tuning of the complex
coupling coefficients ($G$s) in order to ensure the experimentally
observed mass hierarchy \cite{key-34} in the quark sector. These
coefficients remain - as in the SM - free parameters, once the vacuum
expectation values of the scalar field $\phi$ still has to be established.

At this point, one can identify the mass of each quark as \begin{equation}
m(q)=G_{q}\left\langle \phi\right\rangle \label{Eq.40}\end{equation}
 where $q$ in (\ref{Eq.40}) denotes any of the nine quarks in the
model. Note that Eqs. (\ref{Eq.40}) introduce $9$ parameters in
the model.

\subsection{Charged lepton masses}

On the other hand, for charged leptons it was argued \cite{key-12}
that a scalar sextet is a compulsory ingredient in the Yukawa lagrangian
in order to have a realistic and consistent mechanism for generating
masses.

We build this scalar sextet out of the scalar triplets - already existing
in the Higgs sector of the model - as a tensor-like product in the
following manner: \begin{equation}
S=\phi^{-1}\left(\phi^{(1)}\otimes\phi^{(2)}+\phi^{(2)}\otimes\phi^{(1)}\right)\label{Eq.41}\end{equation}
 It plays the same role as the tensor blocks $\chi^{\rho\rho^{\prime}}$
in Eq. (\ref{Eq.3}) . Evidently, $S\sim(\mathbf{1},\mathbf{6},0)$
and thus the generating mass term in the charged leptons sector reads
\begin{equation}
G_{\alpha}{}\bar{f}_{\alpha L}Sf_{\alpha L}^{c}+H.c.\label{Eq.42}\end{equation}
 Hence, consequently the SBB, only positions (12) and (21) in Eq.
(\ref{Eq.41}) will remain non-zero. That is\begin{equation}
\left\langle S\right\rangle =\left(\begin{array}{ccc}
0 & 1 & 0\\
1 & 0 & 0\\
0 & 0 & 0\end{array}\right)\langle\phi\rangle\label{Eq.43}\end{equation}
 The lepton families in the model under consideration here acquire
their masses through the above presented coupling terms (\ref{Eq.41}),
since all couplings due to $S$ get in the unitary gauge the traditional
Yukawa form: $G_{\alpha}\langle\phi\rangle\bar{e}_{\alpha L}e_{\alpha L}^{c}$(according
to a Dirac Lagrangian density put in the pure left form - see Appendix
B in Ref. \cite{key-20}). Therefore, one can identify the mass of
the charged lepton as \begin{equation}
m(e_{\alpha})=G_{\alpha}\left\langle \phi\right\rangle \label{Eq.44}\end{equation}
 Note that Eqs. (\ref{Eq.43}) introduce $3$ more parameters in the
model, in addition to those $9$ necessary ones in the quark sector.

\subsection{Neutrino Mass Matrix }

Since the phenomenon of neutrino oscillations is an undisputable evidence,
all the extensions of the SM must incorporate realistic mechanisms
for generating tiny masses in the neutrino sector of the theory. There
are two main lines in the literature to obtain these tiny masses:
see-saw mechanisms and radiative corrections. For a detailed overview
on theoretical and phenomenological aspects in neutrino physics we
refer the reader to several excellent papers published in the recent
years \cite{key-38} - \cite{key-44}.

We propose here a particular approach that naturally calls for the
canonical see-saw mechanism. The neutrino mass matrix arises from
certain mass terms - regardless they are of the Dirac or Majorana
nature - at tree level in the Yukawa sector of the PPF 3-3-1 model.
This model allows for both kinds of terms, since one can construct
an additional tensor-like product of the form $\phi^{-1}\left(\phi^{(3)}\otimes\phi^{(3)}\right)$
which leads to Majorana mass terms. For Dirac terms one can introduce
terms like $\bar{f}_{\alpha L}\phi^{(3)+}\nu_{\beta R}$. A natural
assumption here is to employ different couplings ($G^{\prime}$s)
in the Dirac sector, while the same parameters involved in the charged
lepton sector ($G$s) are employed in the Majorana sector. This is
quite a natural option, since both the charged lepton masses and the
neutrino Majorana ones are supplied by some tensor-like products of
scalar triplets.

\paragraph{Majorana mass terms}

The model allows for a pure Majorana mass matrix whose elements can
be constructed as a tensor-like product in the manner \begin{equation}
G_{\alpha\beta}{}\bar{f}_{\alpha L}\left[\phi^{-1}\left(\phi^{(3)}\otimes\phi^{(3)}\right)\right]f_{\beta L}^{c}+H.c.\label{Eq.45}\end{equation}
 Such terms develop the well-kown Yukawa shape after the SSB only
in unitary gauge. Therefore one has for the Majorana case - in which
the matrix $M^{M}$ is a symmetric one - the following expression:
\begin{equation}
M^{M}(\nu)=\frac{1}{2}\left(\begin{array}{ccc}
A & D & E\\
D & B & F\\
E & F & C\end{array}\right)\left\langle \phi\right\rangle \label{Eq.46}\end{equation}
 Obviously, the coupling constants are in our notation: $A=G_{ee}$,
$B=G_{\mu\mu}$, $C=G_{\tau\tau}$, $D=G_{e\mu}=G_{\tau e}$, $E=G_{e\tau}=G_{\tau e}$,
$F=G_{\mu\tau}=G_{\tau\mu}$. Moreover, $m(e)=A\left\langle \phi\right\rangle $,
$m(\mu)=B\left\langle \phi\right\rangle $and $m(\tau)=C\left\langle \phi\right\rangle .$

\paragraph{Dirac mass terms}

Assuming the existence of the right-handed neutrinos (see Sec.3.4.2)
one can add to the Yukawa sector terms of the form \begin{equation}
G_{\alpha\beta}^{\prime}\bar{f}_{\alpha L}\phi^{(3)+}\nu_{\beta R}+H.c\label{Eq.47}\end{equation}
 which develop pure Dirac masses. After SSB such a {}``classical''
Yukawa term generates a Dirac neutrino mass matrix: \begin{equation}
M^{D}(\nu)=\left(\begin{array}{ccc}
A^{\prime} & D^{\prime} & E^{\prime}\\
K^{\prime} & B^{\prime} & F^{\prime}\\
L^{\prime} & N^{\prime} & C^{\prime}\end{array}\right)\left\langle \phi\right\rangle \label{Eq.48}\end{equation}
 where primed couplings are self-explanatory.

\paragraph{See-saw mechanism}

With these distinct matrices - Eqs. (\ref{Eq.46}) and (\ref{Eq.48})
- one can construct a canonical see-saw mechanism in order to obtain
the Majorana masses for both the left-handed neutrinos and right-handed
ones. In the flavor basis, the neutrino mass matrix looks like \begin{equation}
M^{D+M}(\nu)=\left(\begin{array}{cc}
M^{M} & M^{D}\\
M^{D} & 0\end{array}\right)\label{Eq.49}\end{equation}
 After its diagonalization, one remains with the two following matrices
assigning for the neutrino masses: \begin{equation}
M^{M}(\nu_{L})\simeq2\left(\begin{array}{ccc}
\frac{\left(A^{\prime}\right)^{2}}{A} & \frac{\left(D^{\prime}\right)^{2}}{D} & \frac{\left(E^{\prime}\right)^{2}}{E}\\
\frac{\left(K^{\prime}\right)^{2}}{D} & \frac{\left(B^{\prime}\right)^{2}}{B} & \frac{\left(F^{\prime}\right)^{2}}{F}\\
\frac{\left(L^{\prime}\right)^{2}}{E} & \frac{\left(N^{\prime}\right)^{2}}{F} & \frac{\left(C^{\prime}\right)^{2}}{C}\end{array}\right)\left\langle \phi\right\rangle \label{Eq.50}\end{equation}
 for the left-handed neutrinos, and \begin{equation}
M^{M}(\nu_{R})\simeq\frac{1}{2}\left(\begin{array}{ccc}
A & D & E\\
D & B & F\\
E & F & C\end{array}\right)\left\langle \phi\right\rangle \label{Eq.51}\end{equation}
 for the right-handed ones.

\paragraph{Neutrino mixing}

The physical neutrino basis can be determined by taking into consideration
neutrino mixing performed by the unitary mixing matrix $U$ ( $U^{+}U=1$).
It switches from the gauge-flavor basis to the physical basis of massive
neutrinos in the manner \begin{equation}
\nu_{\alpha L}(x)=\sum_{i=1}^{3}U_{\alpha i}\nu_{iL}(x)\label{Eq.52}\end{equation}
 where $\alpha=e,\mu,\tau$ (corresponding to neutrino gauge eigenstates),
and $i=1,2,3$ (corresponding to massive physical neutrinos with masses
$m_{i}$). In our case all neutrinos are Majorana fields $\nu_{L}^{c}(x)=\nu_{L}(x)$.
Otherwise, one should consider in the case with neutrinos as Dirac
fields $\nu_{L}^{c}(x)=\nu_{R}(x)$. The mass term corresponding to
neutrino mass yields: \begin{equation}
\mathcal{-L}_{\nu}^{mass}=\frac{1}{2}\bar{\nu}_{\alpha L}M_{\alpha\beta}(\nu)\nu_{\beta L}^{c}+H.c\label{Eq.53}\end{equation}
 The mixing matrix $U$ that diagonalizes the neutrino mass matrix
ensures the relation $U^{T}M(\nu)U=m_{ij}(\nu)\delta_{j}$. It has
in the standard parametrization the form: \begin{equation}
U=\left(\begin{array}{ccc}
c_{2}c_{3} & s_{2}c_{3} & s_{3}e^{-i\delta}\\
-s_{2}c_{1}-c_{2}s_{1}s_{3}e^{i\delta} & c_{1}c_{2}-s_{2}s_{3}s_{1}e^{i\delta} & c_{3}s_{1}\\
s_{2}s_{1}-c_{2}c_{1}s_{3}e^{i\delta} & -s_{1}c_{2}-s_{2}s_{3}c_{1}e^{i\delta} & c_{3}c_{1}\end{array}\right)P\label{Eq.54}\end{equation}
 with $P={\textrm{diag}}\left(e^{i\phi_{1}},e^{i\phi_{2}},1\right)$
- the phase matrix. For, simplicity, we made the substitutions $\sin\theta_{23}=s_{1}$,
$\sin\theta_{12}=s_{2}$, $\sin\theta_{13}=s_{3}$, $\cos\theta_{23}=c_{1}$,
$\cos\theta_{12}=c_{2}$, $\cos\theta_{13}=c_{3}$ for the mixing
angles, and $\delta$ is the CP Dirac phase and $\phi_{1},\phi_{2}$
are Majorana phases. We note here that the later ones can not be removed
by a simple redefinition of the phases, since they carry physical
information for the Majorana neutrinos. They are not active if the
Dirac case is considered.

\paragraph{Mass squared differences}

For physical neutrinos, mass squared differences - which are experimentally
accessible - are defined as $\Delta m_{ij}^{2}=m_{j}^{2}-m_{i}^{2}$.
Their right order of magnitude can be obtained for $\Delta m_{23}^{2}\leq2\cdot10^{-3}$
eV$^{2}$ from Super Kamiokande atmospheric data \cite{key-45,key-46}
and for $\Delta m_{12}^{2}\leq8\cdot10^{-5}$ eV$^{2}$ from solar
and KamLAND data \cite{key-47,key-48} . Considering that in Eq. (\ref{Eq.50}),
the coupling constants act as variables, the diagonalization of the
matrix $M$ is equivalent to a system of $6$ linear equations with
$9$ variables, as $M$ is symmetric, which leads to the following
general solution for the physical neutrino masses: \begin{equation}
m_{i}=F_{i}\left(\frac{\left(A^{\prime}\right)^{2}}{A},\frac{\left(B^{\prime}\right)^{2}}{B},\frac{\left(C^{\prime}\right)^{2}}{C},\theta_{12},\theta_{13},\theta_{23}\right)\left\langle \phi\right\rangle \label{Eq.55}\end{equation}
 where $\theta_{12},\theta_{13},\theta_{23}$ stand for the mixing
angles in the neutrino sector and $i=1,2,3$.

The analytical functions $F_{i}$ could be determined in each particular
case by solving the appropriate set of equations. For the case of
Majorana neutrinos this task was accomplished in the general case
of neutrino mixing without CP-phase violation in Ref. \cite{key-49}.
This case corresponds to the phenomenological situation $\sin\theta_{13}\simeq0$.

The mass squared differences are now: \begin{equation}
\Delta m_{ij}^{2}=(F_{j}^{2}-F_{i}^{2})\left\langle \phi\right\rangle ^{2}\label{Eq.56 }\end{equation}
 With these expressions one can get the mass squared ratio $r_{\Delta}=\Delta m_{12}^{2}/\Delta m_{23}^{2}$
which is independent of the parameters of the scalar sector in the
model - and thus is not affected by the SSB details - and depends
only on the mixing angles and the couplings in the Yukawa sector.

\paragraph{Phenomenological restrictions }

A great deal of experimental data (see \cite{key-44} and Refs therein)
confirm that phenomenological values of neutrino masses $m(\nu_{i})$
are severely limited to a few eVs. Let us compute the sum of the neutrino
masses. It is nothing but the trace of the neutrino mass matrix, \begin{equation}
\sum_{i}m(\nu_{i})={\textrm{Tr}}[M^{M}(\nu_{L})]=2\left(\frac{\left(A^{\prime}\right)^{2}}{A}+\frac{\left(B^{\prime}\right)^{2}}{B}+\frac{\left(C^{\prime}\right)^{2}}{C}\right)\left\langle \phi\right\rangle \label{Eq.57}\end{equation}
 In order to obtain the desired order of magnitude one has to tune
these parameters or even enforce certain symmetries.

\section{Concluding Remarks}

In this paper we have proved that the well-known PPF 3-3-1 model can
be investigated from an exact algebraical viewpoint, by simply using
the method of solving gauge theories with high symmetries proposed
in Ref. \cite{key-20}. In this approach, all the phenomenological
consequences regarding the boson mass spectrum in the model occur
due to a natural tuning of a free parameter $a$. At the same time
the correct couplings of the fermion currents with respect to the
neutral and charged bosons are obtained. We mention that the usual
mixing (small $\phi$ angle) - worked out on the resulting couplings
at the end of the calculus in other papers - is performed in our solution
as an compulsory intermediate step by the method itself. Thus, the
couplings in Table 1 being the exact ones for all the currents in
the model. As one can easily observe, they do not depend on any parameter,
except for the Weinberg angle $\theta_{W}$ well established in the
SM.

A special Yukawa sector is constructed in the fermion sector of the
model in order to generate the correct masses of the particles. Here
a set of $9$ free parameters (Yukawa couplings) are introduced in
the quark sector and $3$ more ones in the charged letpon sector.
As long as the neutrino phenomenology is invoked, one can exploit
it by just tuning $3$ other parameters corresponding to the Yukawa
couplings for the Dirac mass terms, while the same $3$ couplings
from charged lepton sector are employed to ensure the Majorana mass
terms in a suitable see-saw mechanism. Since the unique breaking scale
(with the vacum expectation value $\left\langle \phi\right\rangle $)
is responsible for the necessary SSB, one can establish that for a
$\left\langle \phi\right\rangle $ at around TeV scale, the $A^{\prime}$,
$B^{\prime}$, $C^{\prime}$ have to be in the range $10^{-9}$ in
order to give a viable order of magnitude for the neutrino mass spectrum
$\sum_{i}m(\nu_{i})\sim1$eV.

Our solution presented above offers an exact algebraical framework
for further investigations on interesting topics invoked in some papers
already published on PPF 3-3-1 model. It is able to treat the case
of adding an exotic charged lepton \cite{key-50} which replaces the
right-handed charged lepton in the third position of each lepton triplet.
It can incorporate - if Majorana neutrinos are involved - phenomena
regarding neutrinoless double decay \cite{key-51} and thus lepton
number violation. The particular behaviour of the extra neutral boson
of the theory and its leptophobic character \cite{key-52} - \cite{key-54}
is naturally obtained within our solution. It was argued that such
models can well explain the electric charge quantization \cite{key-55}
- \cite{key-58}. Regarding the neutrino masses, radiative mechanisms
\cite{key-59} - \cite{key-62} could also be employed to generate
tiny masses in contrast to the tree level attempts \cite{key-63,key-64},
while the rich phenomenology of the see-saw mechanism could be further
investigated \cite{key-65} - \cite{key-67}. Anomalous magnetic moment
of the muon \cite{key-68} and other static quantities \cite{key-69}
were calculated using this class of 3-3-1 models. and the perturbative
border (including the Landau pole and the non-perturbative regime)
\cite{key-70} - \cite{key-74} of such models can be also treated
using our elegant parametrization. The search for doubly charged Higgs
bosons \cite{key-75,key-76} can be naturally addressed whithin the
framework of our solution. An attractive possibility stands in exploiting
an additional $U(1)$ symmetry \cite{key-77}.

With such an efficient outcome, we consider that our method acts as
an elegant and viable tool for solving the wide set of theoretical
and phenomenological issues related to 3-3-1 models that, in addition,
could suggest new ways and interpretations for the phenomena already
experimentally confirmed.

\newpage

\appendix

\subsection*{Appendix: Calculating the coupling coefficients}

Our model has three types of fermion triplets. The fundamental irrep
$(\mathbf{3},0)$ of the lepton triplet defines the basic electric
charges of the model, $Q=Q^{(\mathbf{3},0)}(A^{em})={\textrm{diag}}\left(1,-1,0\right)$.
The electric charges in quark's irreps, $(\mathbf{3^{*}},-\frac{1}{3})$
and $(\mathbf{3},+\frac{2}{3})$, are \begin{eqnarray*}
Q^{(\mathbf{3^{*}},-\frac{1}{3})}(A^{em}) & = & {\textrm{diag}}\left(-\frac{4}{3},\frac{2}{3},-\frac{1}{3}\right)\,,\\
Q^{(\mathbf{3},+\frac{2}{3})}(A^{em}) & = & {\textrm{diag}}\left(\frac{5}{3},-\frac{1}{3},\,\frac{2}{3}\right)\,,\end{eqnarray*}
 pointing out the presence of the exotic quarks with the electric
charges $\frac{5}{3}$ and $-\frac{4}{3}$.

The neutral charges of the Weinberg neural boson $Z$ result from
Eq. (\ref{Eq.33}) as \begin{eqnarray*}
Q^{(\mathbf{3},0)}(Z) & = & \frac{1}{\sin2\theta_{W}}{\textrm{diag}}\left(-2\sin^{2}\theta_{W},-1+2\sin^{2}\theta_{W},\,1\right)\,,\\
Q^{(\mathbf{3}^{*},-\frac{1}{3})}(Z) & = & \frac{1}{\sin2\theta_{W}}{\textrm{diag}}\left(\frac{8}{3}\sin^{2}\theta_{W},1-\frac{4}{3}\sin^{2}\theta_{W},-1+\frac{2}{3}\sin^{2}\theta_{W}\right)\,,\\
Q^{(\mathbf{3},\frac{2}{3})}(Z) & = & \frac{1}{\sin2\theta_{W}}{\textrm{diag}}\left(-\frac{10}{3}\sin^{2}\theta_{W},-1+\frac{2}{3}\sin^{2}\theta_{W},1-\frac{4}{3}\sin^{2}\theta_{W}\right)\,,\end{eqnarray*}
 recovering thus all the neutral charges of leptons and standard quarks
predicted by the SM.

The neutral charges of our new neutral boson $Z'$ calculated according
to Eq. (\ref{Eq.34}) read \begin{eqnarray*}
Q^{(\mathbf{3},0)}(Z') & = & \alpha\,(1-4\sin^{2}\theta_{W})\,{\textrm{diag}}\,\left(-2,1,1\right)\,,\\
Q^{(\mathbf{3}^{*},-\frac{1}{3})}(Z') & = & \alpha\,{\textrm{diag}}\,\left(2-10\sin^{2}\theta_{W},-1+2\sin^{2}\theta_{W},-1+2\sin^{2}\theta_{W}\right)\,,\\
Q^{(\mathbf{3},\frac{2}{3})}(Z') & = & \alpha\,{\textrm{diag}}\,\left(-2+12\sin^{2}\theta_{W},1,1\right)\,,\end{eqnarray*}
 where $\alpha=[\sqrt{3}\sin2\theta_{W}\sqrt{1-4\sin^{2}\theta_{W}}]^{-1}$.

For the singlets we obtain simpler formulas since in this case all
the coupling coefficients are given by the chiral hypercharge. Thus
for an arbitrary singlet $({\textbf{1}},y_{ch})$ we have $Q^{(\mathbf{1},y_{ch})}(A^{em})=y_{ch}$,
$Q^{(\mathbf{1},y_{ch})}(Z)=-y_{ch}\tan\theta_{W}$ and $Q^{(\mathbf{1},y_{ch})}(Z')=y_{ch}6\alpha\,\sin^{2}{\theta_{W}}$.

Finally we remind the reader that all the charged bosons have the
same coupling coefficient, $g/\sqrt{2}$, which in units of $e$ reads
$1/\sqrt{2}\sin\theta_{W}$.

\newpage

Table: Coupling coefficients of the neutral currents in PPF 3-3-1
model

\begin{tabular}{cccc}
\hline 
Particle\textbackslash{}Coupling($e/\sin2\theta_{W}$)&
$Z\rightarrow\bar{f}f$&
&
$Z^{\prime}\rightarrow\bar{f}f$\tabularnewline
\hline
\hline 
$e_{L},\mu_{L},\tau_{L}$&
$2\sin^{2}\theta_{W}-1$&
&
$\frac{\sqrt{1-4\sin^{2}\theta_{W}}}{\sqrt{3}}$\tabularnewline
&
&
&
\tabularnewline
$\nu_{eL},\nu_{\mu L},\nu_{\tau L}$&
$1$&
&
$\frac{\sqrt{1-4\sin^{2}\theta_{W}}}{\sqrt{3}}$\tabularnewline
&
&
&
\tabularnewline
$e_{R},\mu_{R},\tau_{R}$&
$2\sin^{2}\theta_{W}$&
&
$\frac{2\sqrt{1-4\sin^{2}\theta_{W}}}{\sqrt{3}}$\tabularnewline
&
&
&
\tabularnewline
$\nu_{eR},\nu_{\mu R},\nu_{\tau R}$&
$0$&
&
$0$\tabularnewline
&
&
&
\tabularnewline
$u_{L},c_{L}$&
$1-\frac{4}{3}\sin^{2}\theta_{W}$&
&
$\frac{-1+2\sin^{2}\theta_{W}}{\sqrt{3}\sqrt{1-4\sin^{2}\theta_{W}}}$\tabularnewline
&
&
&
\tabularnewline
$d_{L},s_{L}$&
$-1+\frac{2}{3}\sin^{2}\theta_{W}$&
&
$\frac{-1+2\sin^{2}\theta_{W}}{\sqrt{3}\sqrt{1-4\sin^{2}\theta_{W}}}$\tabularnewline
&
&
&
\tabularnewline
$t_{L}$&
$1-\frac{4}{3}\sin^{2}\theta_{W}$&
&
$\frac{1}{\sqrt{3}\sqrt{1-4\sin^{2}\theta_{W}}}$\tabularnewline
&
&
&
\tabularnewline
$b_{L}$&
$-1+\frac{2}{3}\sin^{2}\theta_{W}$&
&
$\frac{1}{\sqrt{3}\sqrt{1-4\sin^{2}\theta_{W}}}$\tabularnewline
&
&
&
\tabularnewline
$u_{R},c_{R},t_{R}$&
$-\frac{4}{3}\sin^{2}\theta_{W}$&
&
$\frac{4\sin^{2}\theta_{W}}{\sqrt{3}\sqrt{1-4\sin^{2}\theta_{W}}}$\tabularnewline
&
&
&
\tabularnewline
$d_{R},s_{R},b_{R}$&
$\frac{2}{3}\sin^{2}\theta_{W}$&
&
$-\frac{2\sin^{2}\theta_{W}}{\sqrt{3}\sqrt{1-4\sin^{2}\theta_{W}}}$\tabularnewline
&
&
&
\tabularnewline
$J_{1L},J_{2L}$&
$\frac{8}{3}\sin^{2}\theta_{W}$&
&
$\frac{2\left(1-5\sin^{2}\theta_{W}\right)}{\sqrt{3}\sqrt{1-4\sin^{2}\theta_{W}}}$\tabularnewline
&
&
&
\tabularnewline
$J_{1R},J_{2R}$&
$\frac{8}{3}\sin^{2}\theta_{W}$&
&
$-\frac{8\sin^{2}\theta_{W}}{\sqrt{3}\sqrt{1-4\sin^{2}\theta_{W}}}$\tabularnewline
&
&
&
\tabularnewline
$J_{3L}$&
$-\frac{10}{3}\sin^{2}\theta_{W}$&
&
$-\frac{2\left(1-6\sin^{2}\theta_{W}\right)}{\sqrt{3}\sqrt{1-4\sin^{2}\theta_{W}}}$\tabularnewline
&
&
&
\tabularnewline
$J_{3R}$&
$-\frac{10}{3}\sin^{2}\theta_{W}$&
&
$\frac{10\sin^{2}\theta_{W}}{\sqrt{3}\sqrt{1-4\sin^{2}\theta_{W}}}$\tabularnewline
&
&
&
\tabularnewline
&
&
&
\tabularnewline
\hline 
&
&
&
\tabularnewline
\end{tabular}

\end{document}